\documentclass[sigconf]{acmart}
\AtBeginDocument{%
  }

\usepackage{amsmath}
\usepackage{multirow}
\usepackage{comment}%
\usepackage{float}

\usepackage{dirtytalk}
\usepackage{graphicx}
\usepackage{xcolor}

\newenvironment{sayit}[1]
{\textit{\say{#1}}}

\copyrightyear{2025}
\acmYear{2025}
\setcopyright{cc}
\setcctype{by}
\acmConference[CHI '25]{CHI Conference on Human Factors in Computing Systems}{April 26-May 1, 2025}{Yokohama, Japan}
\acmBooktitle{CHI Conference on Human Factors in Computing Systems (CHI '25), April 26-May 1, 2025, Yokohama, Japan}\acmDOI{10.1145/3706598.3713302}
\acmISBN{979-8-4007-1394-1/25/04}

\begin{document}

\title{The Dilemma of Building Do-It-Yourself (DIY) Solutions for Workplace Accessibility}

\author{Yoonha Cha}
\email{yoonha.cha@uci.edu}
\affiliation{%
  \institution{University of California, Irvine}
  \city{Irvine}
  \state{CA}
  \country{USA}
  }

\author{Victoria Jackson}
\email{vfjackso@uci.edu}
\affiliation{%
  \institution{University of California, Irvine}
  \city{Irvine}
  \state{CA}
  \country{USA}
  }

\author{Karina Kohl}
\email{karina.kohl@inf.ufrgs.br}
\affiliation{%
  \institution{UFRGS}
  \city{Porto Alegre}
  \country{Brazil}
  }

\author{Rafael Prikladnicki}
\email{rafaelp@pucrs.br}
\affiliation{%
  \institution{PUCRS}
  \city{Porto Alegre}
  %\state{CA}
  \country{Brazil}
  }

\author{André van der Hoek}
\email{andre@ics.uci.edu}
\affiliation{%
  \institution{University of California, Irvine}
  \city{Irvine}
  \state{CA}
  \country{USA}
  }

\author{Stacy M. Branham}
\email{sbranham@uci.edu}
\affiliation{%
  \institution{University of California, Irvine}
  \city{Irvine}
  \state{CA}
  \country{USA}
  }




%% Article type: Research, Review, Discussion, Invited or position
%\acmArticleType{Review}
%%
%% Links to code and data
%\acmCodeLink{https://github.com/borisveytsman/acmart}
%\acmDataLink{htps://zenodo.org/link}
%%
%% Authors' contribution
%\acmContributions{BT and GKMT designed the study; LT, VB, and AP conducted the experiments, BR, HC, CP and JS analyzed the results, JPK developed analytical predictions, all authors participated in writing the manuscript.}
%%
%% Sometimes the addresses are too long to fit on the page.  In this
%% case uncomment the lines below and fill them accodingly.
%%
%% \authorsaddresses{Corresponding author: Ben Trovato,
%% \href{mailto:trovato@corporation.com}{trovato@corporation.com};
%% Institute for Clarity in Documentation, P.O. Box 1212, Dublin,
%% Ohio, USA, 43017-6221}
\renewcommand{\shortauthors}{Yoonha Cha et al.}
\renewcommand{\shorttitle}{DIY solutions to support accessible software development}

\begin{abstract}
Existing commercial and in-house software development tools are often inaccessible to Blind and Low Vision Software Professionals (BLVSPs), hindering their participation and career growth at work. Building on existing research on Do-It-Yourself (DIY) Assistive Technologies and customized tools made by programmers, we shed light on the currently unexplored intersection of how DIY tools built and used by BLVSPs support accessible software development. Through semi-structured interviews with 30 BLVSPs, we found that such tools serve many different purposes and are driven by motivations such as desiring to maintain a professional image and a sense of dignity at work. These tools had significant impacts on workplace accessibility and revealed a need for a more centralized community for sharing tools, tips, and tricks. Based on our findings, we introduce the ``Double Hacker Dilemma'' and highlight a need for developing more effective peer and organizational platforms that support DIY tool sharing.
\end{abstract}

\maketitle

\section{Introduction}
Companies in the software development industry have increasingly focused on diversity, equity, and inclusion (DEI) with respect to their workforce. In line with the 154 Fortune 500 companies that published DEI data in 2024~\cite{fortune500_2024}, tech companies such as Google and Microsoft now include disability data as part of their DEI reporting, revealing that 6.7\% and 8.8\% of employees identify as having a disability, respectively~\cite{google_dei_2024, ms_dei_2023}. Actions have been taken to make the software industry more diverse and inclusive, some of these legally binding like the Americans with Disabilities Act (ADA)~\cite{ADA} and the Web Content Accessibility Guidelines (WCAG)~\cite{WCAG} in the United States, the European Accessibility Act~\cite{EAA, de_araujo_quotas_2022} in European countries, and the Quotas Law for People with Disabilities in Brazil~\cite{brazil_quotas, quotas_portuguese}. 

Specifically for Blind and Low Vision Software Professionals (BLVSPs), where there has been an increase in representation year after year~\cite{stackoverflow_general}, the workplace still poses barriers in the form of inaccessible, unusable tools and tasks~\cite{cha_career_2024, cha_participate_2024}. Many tools used for programming, such as Integrated Development Environments (IDEs), have been found to have accessibility bugs~\cite{albusays_eliciting_2016, albusays_interviews_2017, armaly_audiohighlight_2018, armaly_comparison_2018}. Other workplace activities, such as meetings~\cite{cha_participate_2024} and project management~\cite{filho_visual_2015, huff_workexp_2020, pandey_understanding_2021}, are also riddled with accessibility challenges. As a result, BLVSPs may experience reduced career mobility~\cite{cha_career_2024}, in part due to the additional access labor they must perform in the workplace. Often, this labor takes the form of finding and implementing applicable off-the-shelf technologies that may help~\cite{mealin_exploratory_2012}, asking for sighted assistance~\cite{cha_participate_2024, albusays_eliciting_2016, pandey_understanding_2021}, and leveraging ``conversational nudges''~\cite{cha_participate_2024}. However, these strategies are often suboptimal and may only partially mitigate the accessibility problems BLVSPs face at work~\cite{cha_participate_2024}. 

The concept of Do-It-Yourself (DIY) is well known among both software professionals and people with disabilities (PWD), including blind and low vision (BLV) individuals. Software developers are known to create DIY tools in the workplace to save time, help others, or because there is no known solution to the problem they are facing~\cite{smith_diy_2015}. DIY-ing is a necessary practice of many PWD, who are considered the ``original lifehackers"~\cite{Jackson_lifehackers_2018} that spend their lives cultivating intuitive creativity, to navigate a society that is not designed for them~\cite{Jackson_lifehackers_2018}. PWD tend to DIY their own solutions for accessibility (DIY assistive technology), which widely ranges from using existing solutions for different purposes~\cite{hurst_empowering_2011} and combining multiple existing solutions together~\cite{herskovitz_hacking_2023}, to building solutions from scratch~\cite{hurst_making_2013, cha_participate_2024}. While existing research has investigated DIY solutions for accessibility~\cite{herskovitz_hacking_2023, hurst_empowering_2011} and customized DIY tools developed by programmers to meet their needs~\cite{smith_diy_2015, ramler_testtool_2013}, the intersection --- how BLVSPs develop and build DIY tools for accessibility at the workplace --- has yet to be explored. 

To address this gap, we ask the following research questions:
\begin{itemize}
    \item RQ1. What DIY tools do BLVSPs build and use in the workplace, and why?
    \item RQ2. How do these DIY tools impact BLVSPs?
\end{itemize}
We conducted in-depth, semi-structured interviews with 30 BLVSPs who identified as being blind or having low vision, are working or have worked in a software development position either in a corporate setting or as a freelancer, and have at least one year of work experience. Thematic analysis~\cite{braun_using_2006} was utilized for data analysis, and we identified four main themes: (1) key insights about the DIY tools BLVSPs build and use, (2) motivations for building and using DIY tools in the workplace, (3) impacts of DIY tools on BLVSPs' work, and (4) impacts of DIY tools on the entire BLV community. Based on our findings, we discuss the ``Double Hacker Dilemma'' that BLVSPs face: BLVSPs' dual identities of being a life hacker and a software professional (hacker, in the positive sense of the word) creates the dilemma between waiting months or even longer for the company to better support their work situation or taking immediate action because they know how to, yet at the expense of their own additional access labor. Finally, we highlight the need for a community platform for BLVSPs to share DIY tools and knowledge, and offer actionable insights for companies and technology designers. 

Through this research, we contribute the following:
\begin{itemize}
    \item A list of exemplary DIY tools that BLVSPs have built and used in the workplace.
    \item An understanding of BLVSPs' motivations behind building DIY tools and the impact these tools have on their workplace, the BLVSP community, and, in some cases, even the greater BLV community.
    \item An articulation of the ``Double Hacker Dilemma'' that BLVSPs face when they recognize their situation can and should be better, but they either need to wait or do it themselves.
    \item A set of suggestions for companies to better support DIY tool creation and sharing internally, and for the broader BLVSP community to share their DIY innovations and knowledge more broadly.
\end{itemize}

\section{Related Work}

\subsection{Workplace Accessibility of BLVSPs}
Software development requires both social, collaborative work (e.g., meetings~\cite{stray_coordination_2020, meyer_developers_2017, romano_meeting_2001}) and technical work (e.g., programming~\cite{meyer_developers_2017}. Research has identified several challenges faced by BLV workers in the workplace, typically resulting in additional access labor~\cite{das_friends_2019}. The root cause of many of the challenges is inaccessible technologies and tools \cite{das_friends_2019, huff_workexp_2020, mealin_exploratory_2012} and ableist attitudes from colleagues embedded in organizations processes~\cite{cha_career_2024}.
Collectively, the challenges that BLVSPs face in both programming and non-programming tasks hinder their full participation~\cite{cha_participate_2024}, which may hamper their career mobility~\cite{cha_career_2024}. 

Social, collaborative work introduces accessibility barriers for BLVSPs. For example, the unstructured content of \say{stickies} on popular digital whiteboards (e.g., Miro~\cite{miro}) used for collaborative ideation adds difficulty for screen readers to parse content~\cite{dasThatComesHuge2024}, thus BLVSPs have to perform workarounds such as sending ideas to sighted colleagues over chat for them to paste it onto the whiteboards. Project management tools including Jira are also inaccessible for BLVSPs~\cite{filho_visual_2015, huff_workexp_2020, cha_participate_2024}. Although software professionals spend significant time in meetings \cite{meyer_developers_2017}, meeting settings, including when online, create significant hurdles for BLVSPs, leading them to perform significant access labor and in some cases even to stop participation altogether~\cite{branham_invisible_2015, cha_participate_2024, akter_facilitator_2023}.  

Most studies focusing on BLVSPs have investigated the accessibility of technical, programming-related tasks~\cite{mountapmbeme_addressing_2022, albusays_eliciting_2016, armaly_audiohighlight_2018, baker_structjumper_2015, falase_tactile_2019}. Researchers have found that the visually-oriented nature of integrated development environments (IDEs) such as Visual Studio Code~\cite{MS_VSCode} hinder code comprehension~\cite{armaly_comparison_2018}, navigation~\cite{huff_workexp_2020}, and debugging~\cite{albusays_interviews_2017}. For example, glanceability of code~\cite{potluri_CodeTalk_2018} is hindered as screen readers linearly navigate line-by-line through the codebase~\cite{potluri_CodeTalk_2018}. Pair programming also poses accessibility challenges for BLVSPs~\cite{pandey_understanding_2021, huff_workexp_2020, filho_visual_2015}, as AT is usually not installed on their sighted colleagues' machiness~\cite{armaly_audiohighlight_2018, potluri_CodeTalk_2018, potluri_codewalk_2022}. In addition to integrated development environments, BLVSPs commonly use web-browsers and command line interfaces (CLIs) to accomplish tasks (e.g., information seeking, compiling and running code). While command line interfaces are essential for local development and accessing cloud-based services, they pose accessibility challenges for BLVSPs due to the unstructured outputs and poor navigability with a screen 
reader~\cite{sampath_accessibility_2021}.

%One of their daily tasks include seeking information through web searches. To tackle accessibility challenges in information seeking (i.e. inaccessible webpages and documentations), BLV developers implemented workarounds such as custom setups (including scripts) or 

%to the inaccessibility of many sources oFor example, BLV developers found inaccessibility of information seeking sources(e.g. blogposts, documentation sites) As an example, problem solving requiring the seeking of potential solutions on web sites such as blogs, documentation sites, public forums, and general web search \cite{storer_IDE_2021}. Such information seeking comes with its own challenges due to inaccessible web sites. Workarounds included custom setups (including scripts) or asking a colleague for their input  \cite{storer_IDE_2021}. 

%Job specializations within software teams too come with their own challenges. Data scientists use computational notebooks such as Jupyter Notebooks (cite). While accessible interfaces are important, notebook authors can improve the accessibility by improved use of alt text to describe images and the use of data tables to accompany charts \cite{potluriNotablyInaccessibleData2023}. UI development is a 

Researchers have proposed a variety of tools for accessible software development, including replacements for standard tools~\cite{miura_terminal_2024, ulhaque_gridcoding_2022}, command line interface scripts~\cite{ulhaque_gridcoding_2022}, browser plugins~\cite{huh_checker_2024}, and integrated development environment plugins~\cite{potluri_CodeTalk_2018, armaly_audiohighlight_2018}. Some examples include screen reader plugins that help BLVSPs navigate a codebase non-linearly~\cite{albusays_interviews_2017}, a custom application integrated with screen readers to replace the standard command line interface bundled with the Windows operating system \cite{miura_terminal_2024}, a command line interface script to aid debugging code \cite{saben_enabling_2024}, a screen reader friendly Chrome browser plugin that aids BLVSPs in building visual websites \cite{huh_checker_2024}, and a Visual Studio Code plugin to improve collaborative programming between a BLVSP and a sighted professional \cite{potluri_codewalk_2022}. It is unclear how many of these tools are used \say{in the wild}, but their existence shows the demand for improved developer tooling for BLVSPs.

%and CodeTalk \cite{potluri_CodeTalk_2018} that summarizes the information present in the IDE's screen. To address difficulties in skim reading code via a screen reader, an Eclipse plugin (AudioTalk \cite{armaly_audiohighlight_2018}) was developed to provide a better summary more suited for screen readers. Co11ab \cite{dasCo11abAugmentingAccessibility2022a} is a Google docs extension that utilizes audio cues to help non-sighted and sighted colleagues collaboratively author a Google Doc.

% \subsection{DIY Tooling}

\subsection{Maker Culture, DIY-ing, Hacking}

The term ``maker movement'' was originally coined by Dougherty~\cite{dougherty_maker_2012} to refer to people who engage with objects in ways that make them more than just consumers~\cite{dougherty_maker_2012}. Maker culture is driven by a shared belief in the possibilities of creation and innovation through technology, with members as active participants in shaping technology's development and application~\cite{anderson_discovering_2012, lindtner_created_2012, lindtner_hacking_2015, Meissner_diy_2017}. While some view the maker movement as promoting innovation and democratizing technology, research suggests that making is often perceived as a personal lifestyle or leisure activity rather than a political or economic endeavor \cite{Davies_hacking_2018, tanenbaum_democratizing_2013, wang_inventive_2011}.

In 1950, the term \textit{"hack"} was coined at the  Massachusetts Institute of Technology to describe innovative, unconventional approaches to technical issues \cite{Reagle_hackinglife_2019}. Subsequently, the application of "\textit{hacks}" was extended to computing and everyday life \cite{Reagle_hackinglife_2019}. Since then, the idea of "life hacking" has gradually evolved and, today, the term covers the intersection of technology, culture, and larger concerns about work, wealth, health, relationships, and meaning \cite{Reagle_hackinglife_2019}. 

\subsubsection{DIY in the Workplace}
While studies in Human-Computer Interaction (HCI) and Computer-Supported Cooperative Work (CSCW) have increasingly focused on the convergence of DIY, hacking, and craft practices (e.g., \cite{tanenbaum_democratizing_2013, herskovitz_hacking_2023}), literature in this field about hacking in the workplace is sparse. A few articles in the organizational and management literature discuss workplace hacking as a form of ``smart working''~\cite{McEwan_smart_2016} and developing ``shortcuts''~\cite{bloom_hacking_2021} that goes beyond programming tasks. This entails discovering, inventing, and applying tricks for better productivity~\cite{jetha_smarts_2019, stein_leadership_2022} and well-being~\cite{bloom_hacking_2021}, including individual- and team-level hacking activities [94]. Bloom et al.~\cite{bloom_hacking_2021} identify this as another form of labor --- ``working to work'' --- which inadvertently reinforces capitalist systems.

%{[*Some amount of organizational and management literature discuss workplace hacking, describing how the term ``hack'' expanded to be applicable to general work beyond programming to connote ``smart working''~\cite{McEwan_smart_2016}. In the workplace, individuals discover, invent, and apply tricks for better productivity and task management~\cite{jetha_smarts_2019, stein_leadership_2022}, with workplace hacking activities ranging from individual level hacks to team level hacks~\cite{stein_leadership_2022}. In organizational settings, reasons for hacking include satisfying workplace demands as well as their own well-being through applying ``shortcuts''~\cite{bloom_hacking_2021}. Bloom et al.~\cite{bloom_hacking_2021} also refer to ``hacking work" as another form of labor --- ``working to work'' --- undertaken to efficiently manage and complete tasks in the given work time to protect one's well-being which inadvertently reinforces capitalist systems.

Although building tools is considered typical of expert software design work~\cite{petre_software_2016} and recognized as commonly being undertaken by software engineers~\cite{smith_diy_2015}, workplace hacking in the form of DIY-ing tools has rarely been documented in existing literature, with two notable exceptions~\cite{smith_diy_2015, czerwonka2018codeflow}. In a study on the development and use of DIY tools within a company~\cite{smith_diy_2015}, various motivators for software engineers to build DIY tools included desires for efficiency and automation, and because there was no known solution. Some of these tools may be developed clandestinely without the knowledge or approval of a developer's management, and are therefore not tracked in official bug trackers nor stored in official source repositories~\cite{smith_diy_2015}. Other times, the tools are shared across the organization. For example, the tool CodeFlow began as a DIY tool to improve code reviews but is now widely adopted in the organization~\cite{czerwonka2018codeflow}.

%Although DIY-ing software tools in the software development workplace is considered common~\cite{smith_diy_2015} and building a tool is considered a trait of an expert designer \cite{petre_software_2016}, there is little work directly exploring the topic, especially in relation to work hacking. Existing research \cite{smith_diy_2015,ramler_testtool_2013,Allen_barriers_2023} notes that many of the DIY tools in software development come about because developers themselves decide to use their own expertise \cite{petre_software_2016} to build a tool, often to enable them to complete their work more quickly or as part of a side project~\cite{smith_diy_2015}.% 

%Other times, the work is more visible, through a company's \say{inner source} initiative \cite{wan_innersource}, or because developers actively share their tool with colleagues via internal social media channels \cite{smith_diy_2015} leading to wider adoption. For example, the tool CodeFlow began as a DIY tool to improve code reviews but now is widely adopted in the organization~\cite{ramler_testtool_2013}. 

Outside of the workplace, the open-source software movement \cite{bonaccorsi_oss} provides a platform for developers to DIY tools. Open-source software was initially an opportunity for developers to \say{scratch a personal itch} by creating their own projects \cite{raymondCathedralBazaar1999}. An early study identified that key motivators for contributing to open-source included the intellectual stimulation of writing code, and that the code was needed either for work or non-work purposes \cite{lakhani2005hackers}.
%A study on the Linux kernel, noted that one of the key motivators was being identified as a Linux developer \cite{hertel2003motivation} highlighting the community aspect of OSS. 
Moreover, own-use value was categorized as a form of internalized extrinsic motivator in a 2010 literature review that identified ten categories of motivators \cite{von2012carrots}. Altruism was also considered a motivator. Investigation into motivation continued with a more recent study \cite{gerosa2021shifting} identifying that motivating factors had changed since the early days of open-source software, with the importance of own-use declining. 

%Moreover, Von et al. \cite{von2012carrots} reported that own-use value was a form of internalized extrinsic motivator, with altruism and career growth also being motivators. However, recent research \cite{gerosa2021shifting} report a shift in motivational factors for OSS contributors, with personal use becoming less significant compared to earlier studies.

\subsubsection{DIY Assistive Technology}

While specialized assistive technologies (ATs) exist to address different challenges BLV individuals face in life, they often provide generic solutions that do not consider individual differences among users \cite{he_multimodal_2023}. Consequently, BLV individuals have become experts at customizing and "hacking" ATs to suit their unique needs, instead of attempting -- often pointlessly so -- to convince corporations of the needed features and their market viability ~\cite{herskovitz_hacking_2023, Allen_barriers_2023}.

Do-It-Yourself assistive technology (DIY-AT) refers to the development and adaptation of AT devices by non-professionals~\cite{hook_challenges_2014, Bohre_diy_2023}. DIY-AT addresses issues of abandonment and low acceptance rates of commercial ATs by considering user opinions, improving device performance, and adapting to changing needs~\cite{hurst_empowering_2011, Meissner_diy_2017}.
There are different motivations to creating DIY-AT, including increased control over design elements, passion, and cost-effectiveness. The emergence of rapid prototyping tools and online communities has further empowered individuals to create custom AT solutions~\cite{hurst_empowering_2011, buehler_thingverse_2015, Li_perception_2021}. 
Makerspace culture, with its emphasis on bespoke creativity, iterative design, and personal customization, is a promising match for the needs of disabled people, especially for the design of customized assistive technology~\cite{Allen_barriers_2023} at low cost~\cite{hook_challenges_2014, hurst_making_2013, hurst_empowering_2011}. Open design in particular paves the way for the development of customized, affordable ATs~\cite{hamidi_diy_2014}.

%DIY-AT shows promise as it empowers users to create personalized ATs with tailored functionalities at a low cost. The potential of using DIY to create customized digital ATs has long been acknowledged~\cite{hamidi_diy_2014}. 

Some screen readers (e.g., Jaws \cite{jaws} and NVDA \cite{NVDA}) have purposefully been designed to allow the use of plug-ins. %These are 3rd party pieces of software developed by someone other than the software vendor that can be installed by a user to customize the behavior of the screen reader. 
Such plug-ins help to make applications more accessible or enhance the capabilities of the screen reader \cite{momotaz_plugins_2021}. A few of these plug-ins are geared to assisting BLVSPs in coding activities (e.g., \cite{momotaz_usage_2023}). While the plugins are beneficial, challenges remain for BLV(SP)s in identifying suitable plugins due to a lack of a centralized repository, concerns about the security of the plugins, and a lack of financial incentives to develop such plugins \cite{momotaz_usage_2023}.

\subsection{Sharing Knowledge and Tools}

\subsubsection{Sharing among Software Developers}

%The exchange of knowledge among software developers is a critical aspect of the software development process \cite{ebert2008effectively}. %Team size, distribution, and open-source experience influence knowledge-sharing behavior~\cite{Maalej_comprehension_2014}.
To successfully engage communities and to share knowledge requires platforms.
%Community engagement in software development requires strategic communication and knowledge-sharing platforms. 
For example, Google at some point distributed a weekly, one-page printed newsletter to restrooms to increase awareness and adoption of internal tools ~\cite{murphy-hill_toilet_2019}. Within open-source software, an active X (formerly Twitter) community is important as it attracts contributors, impacts popularity, facilitates sharing of resources, and provides a platform for engaging in technical discussions ~\cite{Sharma_twitter_2018, Shimada_sponsors_2022, Fang_twitter_2022, Fan_sponsors_2024}. Mailing lists and social Q\&A communities are also important hubs for project discussions ~\cite{guzzi_msr_2013, vasilescu_social_2014}. Social Q\&A sites in particular function as learning communities and have become valuable repositories of knowledge~\cite{anderson_discovering_2012, guzzi_msr_2013, ahn_learning_2013, vasilescu_social_2014}. 

%The success of knowledge sharing via platforms depends on the ability to provide ways for members to become productive contributors, and play crucial roles in cultivating critical skills~\cite{ahn_learning_2013}. For example, online communities play a crucial role in shaping developers' trust in AI tools through shared experiences and community signals~\cite{cheng_online_2022} and community Q\&A sites can face sustainability challenges due to increasing failure and churn rates, often attributed to low-quality content from undesired user groups ~\cite{srba_stackoverflow_2016}.

%In OSS, the sustained participation of developers can be associated with identity and community construction~\cite{shah_motivation_2006, shah_nature_2004, fang_sustained_2009}. Tweets mentioning GitHub Sponsors' profiles increased sponsor acquisition, funding, and popularity, influencing community involvement and an active Twitter/X community is important to attract contributors because it impacts project popularity, sharing resources, engaging in technical discussions, etc.~\cite{Sharma_twitter_2018, Shimada_sponsors_2022, Fang_twitter_2022, Fan_sponsors_2024}. Mailing lists and social Q\&A communities are also important as hubs for project discussions ~\cite{guzzi_msr_2013, vasilescu_social_2014}, however, social Q\&A sites are shifting support activities away from traditional mailing lists, functioning as learning communities and becoming valuable repositories of knowledge~\cite{anderson_discovering_2012, guzzi_msr_2013, ahn_learning_2013, vasilescu_social_2014}.

\subsubsection{Sharing among PWD}

PWD also use online platforms such as Facebook and X to share knowledge, join communities, and % Such communities enable PWD to 
%share (both provide and receive) information and to 
provide social support to one another, especially to those with a similar disability \cite{sweet_community_2020}. Online communities also provide a venue for PWD to co-create knowledge with researchers and others, leading to product and service improvements that benefit PWD \cite{amann_community_2017}. PWD and non-disabled colleagues are also known to use these communities to engage in joint efforts to create accessible environments, showing interdependent support~\cite{branham_interdependence_2018}.
%sWhen work occurs within shared workspaces, decisions about AT use may be mediated by social interactions, collaboratively, with sighted coworkers \cite{branham_invisible_2015, branham_interdependence_2018}. 
%The creation of online communities often results from the interactions of users through social media, involving various digital communication forms such as text, images, and videos and social networking sites such as Facebook, LinkedIn, and Pinterest; microblogging tools such as Twitter and Tumblr; or media sharing tools such as Instagram and YouTube~\cite{sweet_community_2020}.  Each tool allows users to post information, interact with others, and build online communities according to mutual interests~\cite{sweet_community_2020}.  As a result, the contemporary conceptualization of “community” has evolved from one that necessitated physical proximity to one that includes exchanges among individuals who may never meet face-to-face~\cite{sweet_community_2020}.
Specifically for people within the BLV community with an interest in programming (either as a hobby or profession), the Program-L~\cite{programl} mailing list provides practical assistance on using tools and approaches to aid BLV programmers
%all the while fostering a dynamic community~
\cite{johnson_programl_2022, park_exploring_2023}. %In Program-L, the knowledge diverges between asking and answering questions, and users with high cluster levels in their first year are more likely to interact across roles~\cite{park_exploring_2023}. 
Additionally, an analysis of the use of X reveals an active accessibility community advocating for inclusion and discussing accessibility practices, challenges, and potential solutions \cite{huq_allydev_2023}. 

%Community building for BLV individuals in programming requires fostering a supportive environment, facilitating knowledge sharing, and leveraging diverse communication channels as is seen in Program-L~\cite{programl}, a programming community tailored to BLV individuals with a role in providing support through self-disclosure, practical assistance, and fostering a dynamic community~\cite{johnson_programl_2022}. In Program-L, the knowledge diverges between asking and answering questions, and users with high cluster levels in their first year are more likely to interact across roles~\cite{park_exploring_2023}. From the point of view of social media, Twitter/X conversations reveal to be an active community advocating for inclusion and discussing accessibility practices, challenges, and potential solutions \cite{huq_allydev_2023}. 

\subsection{Research Gap}
While research has examined the development of DIY tools by software developers (primarily within the open-source software community) and also studied DIY-AT tools by PWD, to the best of our knowledge, this study is the first to explore the intersection of work hacking that is realized through DIY tools built by BLVSPs, especially for accessible work. %use at their workplace. 
Our study explores this unexplored area to reveal BLVSPs' experiences with and perspectives on developing DIY tools to tackle workplace inaccessibility.

\section{Methods}
% \usepackage{graphicx}
% \usepackage[normalem]{ulem}
This section describes the recruitment process, participants, interviews, and data analysis undertaken.
\begin{table*}[!ht]

\centering
%\resizebox{\textwidth}{!}{%
\begin{tabular}{c c c c c c c}
\toprule
ID & Job Position & Exp. (yrs)& Org. Type  & Org. Size& Self-Reported Visual Ability & Gender \\ \midrule
P1 & Software Engineer & 6 to 10 & IT  &5,000 +& totally blind & M \\ \hline
P2 & Software Engineer & 1 to 5 & IT  &5,000 +& low vision & M \\ \hline
P3 & Software Engineer & 1 to 5 & Aerospace  &500 to 4,999 & 20/200, limited field vision & M \\ \hline
P4 & Freelancer & 1 to 5 & N/A  &1 to 9 & light perception & M \\ \hline
P5 & Software Engineer & 1 to 5 & Finance  &5,000 +& light perception & M \\ \hline
P6 & Software Engineer & 1 to 5 & IT  &10 to 99& totally blind & NB* \\ \hline
P7 & Accessibility Specialist & 1 to 5 & Finance  &5,000 +& totally blind & M \\ \hline
P8 & Software Engineer & 16+ & IT  &10 to 99& poor orientation & M \\ \hline
P9 & Technical Lead & 6 to 10 & IT  &5,000 +& legally blind & M \\ \hline
P10 & Software Engineer & 11 to 15 & Outsourcing  &100 to 499& totally blind & M \\ \hline
P11 & Accessibility Specialist & 11 to 15 & Non-profit  &100 to 499& legally blind & M \\ \hline
P12 & Software Engineer & 6 to 10 & IT  &5,000 +& totally blind & M \\ \hline
P13 & Freelancer & 1 to 5 & N/A  &500 to 4,999& completely blind & M \\ \hline
P14 & Freelancer & 6 to 10 & N/A  &1 to 9& light / color perception & W \\ \hline
P15 & Software Engineer & 6 to 10 & IT  &10 to 99 & totally blind & M \\ \hline
P16 & Software Engineer & 1 to 5 & IT  &100 to 499 & 2\% useful vision & M \\ \hline
P17 & Accessibility Specialist & 16+ & Education  &5,000 +& totally blind & M \\ \hline
P18 & Software Engineer & 1 to 5 & IT  &5,000 +& low vision & W \\ \hline
P19 & Software Engineer & 6 to 10 & IT  &5,000 +& legally blind w/ tunnel vision & M \\ \hline
P20 & Software Engineer & 1 to 5 & Education  &500 to 4,999& totally blind & M \\ \hline
P21 & Data Scientist & 1 to 5 & Cosmetics  &5,000 +& totally blind & M \\ \hline
P22 & Software Engineer & 16+ & IT  &5,000 +& low vision & M \\ \hline
P23 & Technical Executive & 1 to 5 & Non-profit  &1 to 9 & low vision & W \\ \hline
P24 & Software Engineer & 6 to 10 & Media  &5,000 +& totally blind & M \\ \hline
P25 & Technical Executive & 16+ & Finance  &5,000 +& blind / low vision & M \\ \hline
P26 & Software Engineer & 6 to 10 & IT  &5,000 +& totally blind & M \\ \hline
P27 & Accessibility Specialist & 6 to 10 & IT  &5,000 +& light / color perception & M \\ \hline
P28 & Software Engineer & 1 to 5 & IT  &5,000 +& totally blind & M \\ \hline
P29 & Software Architect & 16+ & IT  &5,000 +& 20/80 corrected & M \\ \hline
P30 & Software Engineer & 16+ & Finance  &100 to 499 & light perception & M \\ \bottomrule
\end{tabular}%
%}
\caption{Detailed information of participants. For participant anonymity, all participant names were replaced with IDs, age is reported in ranges, and job titles do not include specific position information. ``Exp." stands for years of professional experience working in software development. "Org." stands for organization, and the unit of "org. size" column is the number of people. *: NB stands for Non-Binary.}
\label{tab:demographics}
\end{table*}

\subsection{Participants}
We recruited 30 people identifying as being blind or having low vision, who are working or have worked in a software development position (e.g., software engineer, tester, accessibility designer, product manager), either in a corporate setting or as a freelancer, and with at least one year of experience working in the role. %We used these specific eligibility criteria to acknowledge and include the experiences and perspectives of BLV individuals who engage in various aspects of the software development process. 
Participants were recruited through professional contacts, mailing lists such as Program-L (an online discussion group catered to programmers with visual disabilities)~\cite{programl}, and snowball sampling. Most of our participants were located in the United States of America (18), with other participants in Europe (6), India (4), and Brazil (2).

To preserve the confidentiality of our participants, we anonymized identifiable information, including name, age, and job titles. Specific job titles were categorized into generic job categories since titles vary between organizations and thus may be identifiable. For example, Accessibility Specialists include a range of positions such as accessibility tester and consultant. In addition, we report ages in ranges and categorize the type of organization in which our participants most recently worked in the manner of Pandey et al.~\cite{pandey_understanding_2021}. Participant ages ranged from 19 to 59, and years of work experience ranged from 2 to 40 years. We selectively omit participant IDs in places where disclosure may lead to participant deanonymization, and indicate with a footnote. Participants self-reported their vision status: 60\% (n = 18) as being totally or completely blind with little or no light perception, and 40\% (n = 12) a range of visual abilities. Refer to ~\autoref{tab:demographics} for more detailed information on participants.

\subsection{Procedure}
To answer our research questions, we conducted audio-recorded, semi-structured interviews with 30 participants over Zoom between May 2024 and July 2024. The interviews ranged from 43 minutes to 182 minutes with an average of 73 minutes. Prior to the interviews, we emailed participants a study information sheet for review and acquired verbal consent at the beginning of the interview. Participants were also asked to complete a pre-interview survey to collect demographic information. Questions during the interview covered topics such as current accessibility problems participants faced at work and off-the-shelf workarounds they used, followed by questions about tools participants built and their experiences sharing the tools with others. Participants were compensated at a rate of \$40 per hour via either an Amazon gift card or PayPal. This study was approved by the researchers' Institutional Review Board (IRB).

\subsection{Data Analysis}
All 30 audio-recorded interviews were transcribed and checked for quality and accuracy by the researchers. 28 interviews were conducted in English and two in Portuguese (as preferred by these participants) by an author who is a native Portuguese speaker and proficient English speaker. This author translated these two transcripts into English. Thematic analysis~\cite{braun_using_2006} was used for data analysis. The first three authors first analyzed the same seven randomly selected transcripts, performed open coding, and met frequently to identify and refine codes, discuss and refine themes constructed by the researchers, and reach consensus over time. This led to a codebook that the first author more formally documented, which included codes such as ``altruistic motivations'' and ``maintaining professional image.'' After developing the codebook, the first seven transcripts were re-analyzed by the first author, and the remaining 23 were coded individually by two authors, who met frequently to discuss findings, newly identified codes, and possible new or refined themes. When researchers constructed new themes, the authors re-analyzed prior interviews. Throughout the data analysis process, the entire research team met periodically, which led to a refining a few themes and identifying a few new themes. After the first external review of the paper, the authors revisited the themes and combined some lower-level themes for more coherency. The headings and subheadings of the Findings section map to the higher-level themes and detailed themes resulting from our analysis (see \autoref{tab:codes}), with the exception of \autoref{sec:41} where the types of tools used were explicitly identified and categorized.
% \usepackage{graphicx}
%\begin{tiny}

% \usepackage{multirow}
\begin{table*}[ht]
\begin{tabular}{cc}
\hline
High-Level Themes & Themes \\ \hline
\multirow{3}{*}{Motivations for DIY-ing} & Inaccessibility and Indignity of Existing Tools \\
 & Intrinsic and Extrinsic Joys of Hacking \\
 & Efficiency and Autonomy in Relation to Colleagues \\ \hline
\multirow{3}{*}{Impact of DIY Tools on Work} & Increased Accessibility, Confidence, and Equity \\
 & Spending Extra Time Building DIY Tools \\
 & Organizational (Dis)Approval \\ \hline
\multirow{5}{*}{Impact of DIY Tools on the BLV Community} & Sharing DIY Tools \\
 & Collaborating, Advocating, and Offloading \\
 & Uptake of DIY Tools \\
 & Need for Internal Community \\
 & Need for Centralized External Community \\ \hline
\end{tabular}
\caption{The headings and subheadings of the Findings section map to the higher-level themes and detailed themes of our analysis.}
\label{tab:codes}
\end{table*}

\subsection{Researcher Subjectivity, Positionality, and Reflexivity}
The research team consists of researchers in the fields of HCI and Software Engineering, with diverse backgrounds and varying demographics. Researchers are from four different continents, with four researchers being women and two being men. While none of the researchers are BLV themselves, two researchers are highly experienced in accessibility research and engaging with the BLV community; four researchers have extensive experience working in or with software teams in industry; and several researchers have experience in both fields. This diversity of the team enabled the development of the study protocol sensitive to workplace culture and social and technical knowledge. The varied positionalities affected analysis. For example, the accessibility researchers introduced asset-based interpretations~\cite{wong_culture_2020} while the software engineering researchers understood the value of various technical tools in getting work done. To acknowledge and include these varied perspectives, at least three researchers were involved in each data analysis session.

\section{Findings}
%\sayit is italicized \say goes inside \sayit

In each of the below subsections, we address one of the following high-level themes identified in ~\autoref{tab:codes} above.

%We consider DIY tools as any form of technology that has been created by an individual to overcome any specific challenge. Such a tool may benefit other people and so subsequently be shared. This definition is aligned with the concept of DIY-AT discussed by Hurst and Tobias \cite{hurst_empowering_2011}. 
\subsection{DIY Tools Built and Used} \label{sec:41}

As BLVSPs faced accessibility and usability challenges with tools and processes provided by their employers, participants used numerous DIY tools to assist them in the workplace, both for technical and non-technical work. 67\% (n = 20) of our participants explained either building or contributing to DIY tools, while others used existing DIY tools. Tables \ref{tab:tools} and \ref{tab:tools_rest}summarize the 55 different DIY tools that our 30 BLVSPs mentioned, 37 (67.2\%) of which were built by our participants and 18 (32.8\%) were adopted from elsewhere. Adoption rates of the home-built tools varied, with most only used by the author (e.g., \say{AI Content Describer}), some being used by multiple of our participants (e.g., \say{Log Output Accessibilizer}), and others in widespread use in the community (e.g., \say{Can You See Me}). For an extended version of this table, please see supplementary materials.

\begin{table*}[ht]
%\small
\centering
%\resizebox{\textwidth}{!}{%
\begin{tabular}{c c c c}
\toprule
Type & Purpose & Tool Name & Description \\ \midrule
Add-On & T \& I & AI Content Describer & Descriptions for images and UI controls, leveraging multiple GenAI models \\ \hline
Add-On & DS & NVDA Auto-read & Detects new text in the terminal window to notify of changes  \\ \hline
Add-On & PS & Text Information & Provides screen reader in-line definition of selected text  \\ \hline
Add-On & SRO & AccessiNVDA & Provides correct pronunciation of Greek symbols \\ \hline
Add-On & SRO & Speech History & Reviews the last 500 strings of speech synthesizer output \\ \hline
Add-On & PS & Remote Support & Controls a computer running NVDA from another computer running NVDA \\ \hline
Add-On & T \& I & IndentNav & Enables code navigation between different indentation levels \\ \hline
Add-On & SRO & CrashReporter & Attempts recoveries when NVDA crashes, preserving context \\ \hline
Add-On & T \& I & Audio Screen & Enables aural interpretation of images via touch on Windows 8+ screens \\ \hline
Add-On & T \& I & NVDA OCR & Optical Character Recognition (OCR) to extract text from inaccessible objects \\ \hline
Add-On & SRO & Phonetic Punctuation & Converts punctuation and regular expression signs into audio sounds \\ \hline
Add-On & DS & FileZILLA & Labels buttons, adds keyboard shortcuts to FTP (File Transfer Protocol) apps \\ \hline
Add-On & PS & Golden Cursor & Moves mouses with key presses and saves mouse positions for applications \\ \hline
Add-On & T \& I & SVG to PNG & Converts images in SVG format to PNG format to allow NVDA access \\ \hline
Script & DS & InclusiAI & Transparent window reading inaccessible terminal outputs for screen reader \\ \hline
Script & DS & MongoDB Navigator & Allows easier navigation and editing of MongoDB data on notepad \\ \hline
Script & DS & Visual Studio Autocomplete & Allows autocomplete pop-ups on Visual Studio with screen readers \\ \hline
Script & DS & CLI Pathname Simplifier & Reads out only the last backslashed path from an entire path in the CLI \\ \hline
Script & T \& I & JSON Viewer & CLI to read websites, transform to JSON, and see content in a tree view \\ \hline
Script & DS & GitHub Project Manager & Clones Github repositories and enables operations with fewer key presses \\ \hline
Script & PS & Chromium Operations & Launches website and performs operations using a headless browser \\ \hline
Script & DS & Code Review System & CLI that identifies code changes in the code review system \\ \hline
Script & T \& I & Cleaner for Swagger & Cleans C\# code generated by Swagger, for intuitive use with screen reader \\ \hline
Script & DS & Log Output Accessibilizer & Parses output of code execution, reading necessary lines only \\ \hline
Script & DS & Accessible Code Signer & Makes it easier to use a code signing certificate to sign exe and dll files \\ \hline
Script & PS & XBindKeys & Binds commands to certain keys or key combinations \\ \hline
Script & DS & Notifications Monitor & Manages notifications in Linux distributions and tells user \\ \hline
Script & CS & PPTX2MD & Converts slide decks into screen reader friendly Markdowns \\ \hline
Script & DS & Dashboard Accessibilizer & Retrieves visual information in dashboards through API calls \\ \hline
Script & T \& I & Monitor and Read & Takes partial screenshot of screen and OCRs only that part \\ \hline
Script & CS & Volume Docker & Decreases meeting volume when screen reader speaks \\ \hline
Script & DS & SQLite Preview Navigator & Quickly navigates to SQLite preview with screen reader \\ \hline
Script & PS & Data Monitoring System & Pulls data from company products and presents it in an accessible way \\ \hline
Script & T \& I & Statistics Retrieval System & Directly pulls data from API to bypass inaccessible internal statistics system \\ \hline
App & CS & Can You See Me & Offers guidance on position within camera for independent positioning \\ \hline
App & PS & Win My AI & Computer-Android emulator bridge to run Be My Eyes on Windows Desktop \\ \hline
App & DS & GPTCMD & Command line interface wrapper tool to help interface with OpenAI models \\ \hline
App & DS & Text-To-Speech in C\# & TTS for Linux, Android, and Javascript which works without internet \\ \hline
App & PS & HTML Accessibilizer & Converts inaccessible PDF including formulas, etc. to an accessible HTML \\ \hline
App & DS & Android Client & TCP socket to make phone calls and access texts through a computer \\ \hline
App & T \& I & VOCR & AI-powered OCR that integrates with VoiceOver \\ \hline
App & DS & SQL Manager Lite & Accessible, light version of Microsoft SQL Manager \\ \hline
App & DS & Log Filtering Tool & Pastes log output into text editor and filters out logs according to need \\ \hline
App & PS & Graphics Color Inverter & Inverts the color scheme only on the local machine on the graphics level \\ \hline
App & DS & Hex Editor & Screen reader accessible hex editor \\ \hline
App & PS & Brightness App & Sets display brightness either automatically or to predefined values \\ \hline
App & PS & Magnifier Glass & Enables zooming in on parts of the display on Windows 10 \\ \hline
Plugin & T \& I & Button Labeler & Javascript extension to label unlabeled buttons \\ 
 \bottomrule
\end{tabular}
%}
\caption{DIY tools that BLVSPs built or used. The authors of this paper assigned temporary names to some tools that did not have official names for reference purposes. ``Add-On" signifies an NVDA add-on.}
\label{tab:tools}
\end{table*}

\begin{table*}[ht]

\centering
%\resizebox{\textwidth}{!}{%
\begin{tabular}
%{p{0.05\linewidth}<{\centering}
%p{0.1\linewidth}<{\centering}
%p{0.15\linewidth}<{\centering}
%p{0.6\linewidth}<{\centering}}
{c c c c }
\toprule
Type & Purpose & Tool Name & Description \\ \midrule
Plugin & PS & Accessibility Agent & Internal AI chat tool searching accessibility documents and provides information \\ \hline
Library & T \& I & Web Accessibilizer & Web framework to help make HTML pages more accessible \\ \hline
Library & T \& I & MathCat & Converts MathML to speech and braille \\ \hline
Screen Reader & T \& I & Eloud & Less verbose screen reader for Emacs using a speech synthesizer \\ \hline
Screen Reader & T \& I & Chatterbox & Text-to-speech solution in Swedish or English, for small pieces of text \\ \hline
Screen Reader & T \& I & TDSR & Two Day Screen Reader. Command line interface screen reader for MacOS and Linux \\ \hline
Linux Distribution & DS & TalkingArch & Accessible version of Arch Linux live ISO image  \\ \bottomrule
\end{tabular}
%}
\caption{DIY tools that BLVSPs built or used (continued from \autoref{tab:tools}).}
\label{tab:tools_rest}
\end{table*}

\subsubsection{Purposes of DIY Tools}

We categorized the DIY tools at the workplace as follows:
\begin{itemize}
    \item \textbf{Text/Image Interpretation Support (T \& I):} These tools helped BLVSPs interpret inaccessible information. Some helped describe images (e.g., \say{AI Content Describer}, \say{VOCR}, \say{Monitor and Read}), others helped to better present code through screen readers (e.g., \say{IndentNav}, \say{Cleaner for Swagger}), yet others  were built to solve inaccessibility in internal corporate tools (e.g., \say{Statistics Retrieval}).
    \item \textbf{Collaboration Support (CS):} These tools were designed to aid collaborative tasks such as meetings and presentations (e.g., \say{Can You See Me}, \say{Volume Docker}, \say{PPTX2MD}).
    \item \textbf{Development Support (DS):} These tools were designed to make specific software development tasks more accessible. \say{SQLite Preview Navigator}, \say{SQL Manager Lite}, and \say{MongoDB Navigator}, for instance, support easier access to specific databases. As two other examples, \say{Log Filtering Tool} and \say{NVDA Auto-read} identify changes in files and monitor log files in an accessible manner, respectively.
    \item \textbf{Productivity Support (PS):} These tools boosted BLVSPs' productivity by making keyboard usage more efficient with the screen reader (e.g., \say{Golden Cursor}).
    \item \textbf{Screen Reader Optimization Support (SRO):} These tools facilitated better interaction of participants with their screen readers. For example, \say{AccessiNVDA} provides accurate pronunciation of Greek symbols that used to be incorrectly pronounced.
\end{itemize}

Notably, only a handful of the DIY tools were company specific, for instance to access otherwise inaccessible internal company data (e.g., \say{Dashboard Accessibilizer}). The vast majority, however, would be well-suited for any workplace environment. A subset of these tools, especially several related to text/image interpretation support and collaboration support, are helpful beyond BLVSPs to BLV information workers more generally. Finally, many of the DIY tools were built and used to assist BLVSPs in programming, highlighting the essential role DIY tools have in supporting accessible software development.

Some DIY tools started off as a solution to address an issue in a developer's personal life but later became essential for work. P7 was unable to read memes sent by their friends, so they developed a tool which they realized could help them in the workplace: \sayit{So after I wrote it... it spiraled into... a transformative type of technology that we really, really, desperately need in not just my industry, but any blind or visually impaired individual who uses the computer needs something like this.} Companies thus benefit from developers' willingness to tinker in their spare time.

Interestingly, in compiling \autoref{tab:tools}, we found duplicate effort put into some of the DIY tools built by our participants. \say{Dashboard Accessibilizer}, \say{Log Output Accessibilizer} , and \say{CLI Pathname Simplifier} were built by a different participant each, yet effectively performed the same function, sometimes for the identical software. It appears that our participants were either unaware of the existence of similar tools or decided to build their own regardless. 

\subsubsection{Types of Tools Built/Used}
BLVSPs were flexible in the types of DIY tools they built or adopted, as the technology---ranging anywhere from small scripts to standalone, interactive applications---\sayit{depends on the task} (P12).

Among the 55 DIY tools built or used by BLVSPs, the most common type of tool was scripts (n = 20) followed by NVDA add-ons (n = 14), and applications (n = 13). 

Many participants described the development of DIY CLI scripts (n = 20). P16 noted having developed as many as 132 scripts (only some of which we touched upon in the interview; those are reflected in the \autoref{tab:tools}). The complexity of scripts ranged widely, from just simplifying pathnames to leveraging available Application Programming Interfaces (APIs) to build the textual equivalent of otherwise graphical standard user interfaces (UIs). P12 built a script that enabled them to access information previously inaccessible on a corporate dashboard: \sayit{Anything you can do using a visual UI, you could essentially do using a bunch of APIs as well. For me it was basically making those thread calls to specific actions, whether a \say{get} or \say{post}, and getting kind of those things done.} 

For the next most commonly mentioned tools, NVDA add-ons, participants appreciated that NVDA allows writing add-ons: \sayit{The fact that NVDA has add-ons, and the fact that there are people with skills who have shared them [DIY tools], it's pretty nice} (P15). The flexible architectures of browsers were equally leveraged by some BLVSPs to develop plugins to make platforms (e.g., websites) accessible, which many participants reported frequently doing so: \sayit{Usually the problem with these... is the the buttons are not labeled... So, I have them labeled through JavaScript and stuff. I do that, a lot.}

Less frequently developed were screen readers (n = 3), libraries to be incorporated into applications (n = 2), and an accessible version of the Linux Distribution (\say{TalkingArch}). The reduced number perhaps reflects the complexity of these more full-fledged endeavors in comparison to the typically easier effort of developing plug-ins and scripts, as P24 commented: \sayit{It's worth drawing a distinction between tools and scripts, right? So when I say a \say{script,} I mean something that isn't as fully featured.}

\subsection{Motivations for DIY-ing}
Our participants had a wide range of motivations for building and using DIY tools at the workplace. 

\subsubsection{Inaccessibility and Indignity of Existing Tools} \label{sec:421}
BLVSPs were motivated to build and use DIY tools because companies failed to provide adequate accommodations, and suboptimal workarounds were indignifying. 
%\paragraph{Company's Failure to Provide Reasonable Accommodations}
Taking matters into their own hands was an alternative to submitting tickets for accessibility troubleshooting and waiting indefinitely. P7 expressed bitterness about the \sayit{sad state} of accessibility request turnaround times. Although he reached out to a vendor \sayit{months, and months, and months ago,} asking \sayit{Hey, can you just make this one code change?}, he never heard back, concluding that
%: \sayit{I'm still waiting on the response... That's where it dies, usually.} 
\sayit{sitting on a long-haul flight and coding out a solution} was easier. 
%\sayit{It's easier to write a modification like that than it is to reach out to the company, and advocate, and say, \say{Hey, can you just make this one code change?}} 
This sense that companies simply would not do the work was pervasive. For example, P22 stated \sayit{I just don't think anyone is going to do it for us,} and P24 explained, \sayit{[my need] was never going to be solved by a mainstream company, so I rolled up my sleeves and did it.}
Even when BLVSPs proactively wrote the code fix and sent it to the company, the issue remained unsolved: \sayit{It kind of gets to me on a primal level, like if I'm sending you the code to fix it... It's still going to sit in the backlog forever and ever, and ever} (P7).
%\paragraph{The Stolen Dignity by Inadequate Accommodations}

Some participants chose to try various off-the-shelf workarounds before resorting to DIY-ing. However, these were often suboptimal and ultimately drove participants to create their own solutions. P22,  tired of struggling with unreliable tools, shared: \sayit{It is awful, it steals my dignity and makes me feel like an oaf... For a while I struggled with it. I was like, \say{Is it me? It's not me. It's the tools. I'm going to build the tools. I'm perfectly capable.}}

%\paragraph{no one else will do it, I have to, no choice.. which is usually not a tenet of DIY}} 
The involuntary DIY-ing we see here contradicts the mainstream notion of DIY-ing as a voluntary, recreational activity.

\subsubsection{Intrinsic and Extrinsic Joys of Hacking}
Our participants identified hacking as a result of being intrinsic problem solvers and eager to help fellow BLV individuals with their DIY tools.
%\paragraph{Being an Intrinsic Problem Solver}

Several participants described themselves as intrinsic problem solvers who happened to have the domain expertise in software development, which motivated them to build DIY tools. Some BLVSPs believed that their \sayit{brain is probably wired that way} (P8), saying: \sayit{I want to resolve this problem. Sometimes it's not worth the time, but often, it's fun} (P8). P24 also described that \sayit{[DIY-ing]'s happened quite a few times now at work,} as his life being a blind person has been \sayit{one massive life hack all the way through, constantly having to come up with workarounds from a really, really young age.}
%\paragraph{Altruistic Motivations}

Participants also had altruistic motivations as members of the BLV community, namely to help other BLV individuals. P15 described his three-fold motivation: \sayit{It's part tinkering, part trying to help other people, and part just because I can.} P13 shared the satisfaction that comes from helping other people:  \sayit{it's also very satisfying when other people also find it helpful.} Similarly, P24 shared: \sayit{I like the idea of putting stuff out there to help people}.

We see how building and sharing DIY tools are activities motivated by one's software professional identity as well as one's blindness identity.

%\paragraph{introduce the double hacker term here...?}
%In our participants' intrinsic and extrinsic motivations for building DIY solutions, we observed BLVSPs' double hacker identities: (1) the software professional hacker identity and (2) the BLV hacker identity. The former identity led to problem-solving by programming DIY solutions based on their expertise. As members of the BLV community, BLVSPs' latter identity resulted in external motivations to help out other BLV individuals facing similar problems.

\subsubsection{Efficiency and Autonomy in Relation to Colleagues}
Participants desired to achieve higher efficiency and autonomy to maintain a professional image through DIY tools, which sometimes was not feasible with existing workarounds such as off-the-shelf technology and sighted assistance.
%\paragraph{Wanting to be Faster at Work}
%Our participants built and used DIY tools to increase their work efficiency, wanting to avoid the additional labor that comes with being a screen reader user, and to stay apace with sighted colleagues. %

Just like any other software professional, our participants leveraged their domain expertise to get work done more efficiently: \sayit{I'm not good at Excel. So if someone gives me a CSV... if I need to do any kind of analysis on it... my perception is that, \say{You know what? It'll be quicker to just write some quick and dirty code to do it}} (P24). Participants also referred to their \sayit{laziness} (P15): \sayit{The stuff was accessible anyway, but instead of it being a six-step process, it's a one-step process because I wrote an add-on for something to do it} (P28). 

Unlike sighted software professionals, however, BLVSPs had an additional need to be efficient in the face of accessibility and usability barriers. 
When software is technically accessible but is not designed for screen reader use, the interaction becomes tedious and slow--a phenomenon we refer to as the ``accessibility tax." What is often one click away for a sighted professional involves many key presses for BLVSPs. %where \sayit{with all that crap, it's harder to navigate} (P8). 
P16 described: 
\begin{quote}
    \sayit{I guess as a good blind person, you're supposed to start hitting the tab key 4,000 times. And then hitting enter when you hit the right thing... and it's only because you're forced to use this sighted paradigm. If you were going to build a blind-first interface, it wouldn't be a GUI [Graphical User Interface].} 
\end{quote}
To circumvent this, P16 built non-GUI interfaces to \sayit{just interact with the API} and avoid the tax. Similarly, P8 described building his own tool to navigate JSON files, which \sayit{washed away the unnecessary crap, and placed it in a tree view, so you can expand and collapse. So it was much easier to navigate the JSON output [with fewer keystrokes].} 

Participants were particularly concerned about how the accessibility tax would affect their professional image. P24 explained: \sayit{I care a lot about the pace that I can do things at. It's not to do with my technical knowledge, but I am slower at a lot of things. And that can have a bit of an effect in terms of, \say{Well, are people going to want to pair with me?}} P22 shared his frustration when he lost access to tools after losing his vision mid-career: 
\begin{quote}
    \sayit{It really hurts me that I don't have the tools to [work fast] anymore. I also feel wasted, because I know what I can do. Other people don't. What they see when I use the tools that are available is someone who's fumbling around and struggling... And that feedback causes me crazy anxiety. It is horrifying... I hate that.}
    %I'm tired of it.}
\end{quote}
Through creating their own tools, participants sought to \sayit{have the same speed as everybody else} (P1) and \sayit{compete effectively, and be someone who's valuable} (P22).

%\paragraph{Wanting to be Independent and Autonomous}
%Participants did not want their professional impressions to be damaged by inaccessible tools at work. 
In some cases, BLVSPs responded to the accessibility tax by \sayit{fall[ing] back on sighted assistance} (P28) from colleagues.
Participants described that %they \sayit{won't continue to waste time} (P27) trying to solve a problem after a certain point, as 
\sayit{sometimes, using five different tech options is not an efficient use of time} (P27), and \sayit{the best option is just getting someone sighted to help you out} (P12). Similarly, P9 found asking for sighted assistance \sayit{a pain, but it works.}

However, many others worried about negative implications sighted assistance could have on their careers as well as their colleagues'. P24 had to ask \sayit{whether I was in shot of my webcam or not} in every meeting, which made it difficult to maintain a professional image: \sayit{a sighted person might think, `Oh, wow! They can't even tell whether his face is in shot of his webcam or not. That's really, really easy! If he can't do that, then what else can't he do?'} %Although this bore no relation to his software engineering skills, he was \sayit{worried about what my asking was doing to people's perceptions of my abilities,} which would impact his professional image.
Relatedly, P6 was concerned colleagues would become resentful: \sayit{Asking for other people's time, and potentially having them delay their work might show adverse effects on [their colleagues' careers]. So that is one of the scary parts about reaching out and getting assistance.} 

For these participants, their motivation to DIY-ing was \sayit{primarily personal autonomy and independence. Not having to depend on sighted people for me to do my work} (P12). DIY-ing was particularly appealing for recurring tasks: \sayit{If it's something that I need to do very routinely, maybe more than once every couple months, I like to create a process for it, so I don't have to ask people} (P7).

BLVSPs were faced with two unappealing options: risking peers' disrespect and resentment by seeking assistance, or risking low work productivity by declining it. DIY-ing became the only way out.

%Participants desired to achieve personal autonomy and independence through DIY tools, which sometimes was not feasible with existing workarounds such as off-the-shelf technology and sighted assistance.

%4.3.1+preamble as of 9/11 11pm: 501 words
\subsection{Impact of DIY Tools on Work}
The DIY tools that BLVSPs built and used enhanced workplace accessibility, confidence, and equity, but such benefits were counterbalanced by significant access labor and friction with organizational policies.

\subsubsection{Increased Accessibility, Confidence, and Equity}
%Participants shared how DIY tools \sayit{absolutely increased the quality of my work life} (P12). They felt more productive, confident, and on equal footing with sighted colleagues, and sometimes even got promotions. 

%\paragraph{Access to the Software Development Profession}
As described above, BLVSPs were motivated to DIY to increase accessibility, restore dignity, and enhance independence and efficiency; in large part, their tools delivered. 
In some cases the DIY tool was so critical that, without it, the BLVSP may not have been able to continue working in the field. P16 recounted the anxiety of setting up an accessible Linux environment upon starting a new position, noting that failure would have rendered him \sayit{unable to start the job.} Similarly, P22 reported: \sayit{I would not have been able to continue programming [after losing my vision] without something like Eloud.} 
An important theme in this regard is independence: the use of DIY tools did not just make them able to work as software professionals, but on many an occasion enabled them to perform work on their own. 
% this paragraph feels too choppy 
%BLVSPs also appreciated that DIY tools increased their independence. 
For example, P7 was able to independently perform accessibility testing with a DIY tool, whereas previously he \sayit{had to always ask my colleagues} to verify aspects such as zooming, contrast, and underlined links.

%\paragraph{Staying Apace of Sighted Colleagues}
With DIY tools, increased independence translated to increased speed and parity with sighted colleagues. For example, P26 credited \say{PPTX2MD} for \sayit{drastically increas[ing] my speed} in reviewing and preparing accessible software development presentations, making it \sayit{invaluable to my time, and everyone else's time.} %Despite occasional issues such as having to \sayit{take 90 minutes out of the middle of the workday and fix the stupid bug [of DIY tool]} (P16), participants found that the maintenance was worth the investment.
%Without DIY tools, participants \sayit{find that my productivity is cut} (P7).
As a result, building DIY tools, \sayit{level[s] the playing field inasmuch as it might make me quicker at my job} (P24). In fact, DIY tools sometimes allowed BLVSPs to surpass their sighted colleagues. The set of strategies and technologies P17 developed for work \sayit{made me actually more efficient than my sighted counterparts,} as the DIY tools allowed him to bypass the traditional GUI paradigm, or \sayit{the way that sighted people did it.}

%\paragraph{Increased Autonomy and Confidence}
%BLVSPs also appreciated that DIY tools increased their independence. For example, P7 was able to independently perform accessibility testing with a DIY tool, whereas previously he \sayit{had to always ask my colleagues} to verify aspects such as zooming, contrast, and underlined links.

As a result, BLVSPs became more confident at work, especially in the presence of sighted colleagues. P24 felt \sayit{a lot more relaxed} during videoconferencing meetings, as \say{Can You See Me} ensured he was properly \sayit{in shot}. For P22, the additional streamlining offered by DIY tools was key: \sayit{I feel like I'm finally able to use a computer with dignity again... I don't have to do all these stupid workarounds while people are looking over my shoulder going, \say{What the fuck is he doing?}}

% this doesn't really relate to BLVSPs work, but maybe we can integrate some into Getting Positive Unexpected Results

%P17 argued, \sayit{If you can think creatively, you can be more effective.} 

%\paragraph{Unexpected Accolades}
Sometimes, BLVSPs' efforts were rewarded with unexpected accolades. One participant{\footnote{We omit specific details in this example to reduce risk of deanonymization.}} wrote DIY software to replace an inaccessible video playback system, which \sayit{actually worked really well for the rest of the team as well.} The team \sayit{ended up shelving the commercial tool,} and he got promoted from an associate to a junior software engineer. Another BLVSP\textsuperscript{1} was acknowledged as an expert on a specific API after using the API to create a DIY workaround to an inaccessible analytics system: \sayit{Every now and then, [coworkers' response to DIY-ing] does surprise me, and something good will come from it.}

Overall, many participants voiced that DIY tools \sayit{absolutely increased the quality of my work life} (P12). They felt more productive, confident, and on equal footing with sighted colleagues, and sometimes even got promotions.

\subsubsection{Spending Extra Time Building DIY Tools} 
While some BLVSPs received company support to build DIY accessibility tools during work hours, many spent personal time \sayit{engineering my way around accessibility} (P24). Some participants resented using personal time, some accepted it, and some enjoyed or learned to enjoy this process.

%\paragraph{DIY-ing on Company Time}
Some participants were allowed to build DIY tools during their work hours. P12 praised his company for letting him \sayit{spend three weeks in a year designing and developing for accessibility} that he can use \sayit{for the remaining 49 weeks.} Others negotiated development time on a per-tool basis by reframing accessibility problems as usability problems that \sayit{benefit the whole team if it's fixed}:
\begin{quote}
    \sayit{If you phrase it as just an accessibility problem, then realistically, you're either doing that in your own time, or during lunch... If you take the problem and rephrase it,... you can usually find the time to do it in the BAU [Business As Usual] time.} (P24)
\end{quote}

%\paragraph{DIY-ing on Personal Time} \label{sec:432}
More frequently, however, BLVSPs built DIY tools \sayit{in [their] personal time} (P25), although \sayit{it's not really something that I should have to do outside of company hours} (P12). The time spent building their own solutions ranged anywhere from 30 minutes to several sleepless week nights and unrestful weekends. P24 shared how \sayit{it's affecting my career negatively}:
\begin{quote}
    \sayit{Sometimes I've had to really draw on my energy reserves... As a blind software engineer, you might have Monday, where literally everything that you do just doesn't work, because it's all inaccessible. And then, you have to spend your evening engineering your way around these accessibility problems, and you can't go to bed until you've got that done. Not only that, but you have to have enough energy to go in on Tuesday, with a smile on your face, as if nothing's happened. Everyone else got a good night's sleep, but I only got two hours, right?}
\end{quote}
%This may be perhaps because BLVSPs weren't able to find time within work hours: \begin{quote} \sayit{The problem is, when do I develop these solutions? Because I've got ideas for tools that I could write to sort of speed this up for me, but there's never the time. And it's like, \say{Well, do I not take lunch? Do I start work late? Do I finish work early [to build DIY solutions]?}} (P24) \end{quote}

%\paragraph{Feeling Resentment}
As a result, many participants expressed resentment towards having to build DIY tools outside of work hours. BLVSPs felt strongly against being \sayit{used as beta testers} (P25), and that \sayit{blind people should not be used as solutions providers for products at work that should be accessible} (P25). The time spent on DIY accessibility workarounds infringed on their ability to \sayit{take a break from being blind} (P16) and spend time \sayit{work[ing] on other interesting [software] problems} (P16) that are \sayit{purely for myself} (P13). Even P24, who earned a promotion based on his DIY work, did not perceive DIY-ing as career development and was keenly aware of the inequity: \sayit{You could argue that working on accessibility software is career development. Because it's going to enable me to progress. But I just don't see it like that... because taking care of accessibility is not my job. And no one else has to do it.}

%\paragraph{Feeling Resigned}
Several BLVSPs resignedly accepted the necessity of DIY development effort: \sayit{it's annoying... But what's the alternative?} (P24). For P7, it was simply \sayit{inevitable}, as he described: \sayit{It's not ideal. But I don't know another way... as a blind individual, there's extra work that you're going to have to put in.}
Similarly, P13 expressed reluctant acceptance: \sayit{Now, it's just life. It's less annoying now, just part of work, even though it's not part of work.}

%Similarly, DIY effort at work was expected by P25, because \sayit{the world is not built to accommodate us [PWD]}. 

%\paragraph{Finding Silver Linings}
Despite feeling resentment and resignation, some participants enjoyed DIY-ing or tried to focus on its positive aspects.

Those who enjoyed DIY-ing alluded to their love of problem solving. For P25, it is a \sayit{pretty clear win-win} since \sayit{I get to write code, and I get to make things easier for myself.} Yet, for others, finding enjoyment required a bit more effort: \sayit{I'd much rather not have to... But there's almost always a silver lining} (P24). P13 explained his coping mechanism of \sayit{trying to figure out how I could enjoy this more and think of it less as work}, identifying DIY solutions at the intersection of what he \textit{wanted} to do and what \textit{needs} to be done. Ultimately, feelings of resentment tended to overshadow enjoyment. P22 believed that there was a \sayit{double-edged sword to that extra work that blind people have to do... Work outside of work that benefits work.} Similarly, reflecting on his experiences, P24 shared: \sayit{There are good points around having to [DIY], but to be clear, I don't think they make up for the negatives.}

\subsubsection{Organizational (Dis)Approval}
Organizations had different policies on the use of DIY tools and open source assistive technologies, which varied widely depending on the company's size and flexibility. Restrictions compelled BLVSPs to keep DIY tools hidden to maintain their productivity and job security.

%\paragraph{Smaller versus Larger Companies} \label{sec:433}
Some organizations---mostly smaller-sized companies---took a flexible approach to permitting DIY tools. For instance, one participant noted, \sayit{it's a 200-person company, things are more relaxed and they don't enforce very strong policies on us} (P10). Conversely, larger corporations imposed stricter regulations and required excessively complex approval processes, which were cumbersome and time-consuming. One participant\textsuperscript{1} who had worked for both a startup and a large corporation explained the dichotomy: \sayit{It's kind of night and day [between two organizations]... The smaller company doesn't care, as long as you don't leak important data,} while the large company enforced \sayit{a lot of policies. We are bound by so many different restrictions.} This contrast is well illustrated by the usage of DIY tool \sayit{AI Content Describer}; while the small companies that employ P10 and P15 permitted its use, the larger companies that employed P24 and P25 did not. 

%In such companies, so long as BLVSPs were mindful of confidentiality, there were no issues \sayit{if we're not dropping some kind of internal confidential information, they won't make an issue of it} (P15).

%\paragraph{Laborious Approval Processes} \label{laborious_approval}
%labor involved in review
Companies' approval processes to review BLVSPs' DIY tools and authorize them for use at work were lengthy, required additional access labor, and were unlikely to return a positive result. P7 described needing to weigh productivity against compliance: \sayit{You have two options: do it and shut up about it... or request permission... and wait a few months for them to maybe get back to you.} Even after months of waiting, participants faced the extra labor of educating the compliance team about accessibility, only to be denied use of the DIY tool: 
\begin{quote}
    \sayit{Someone gets back to you, and has, [with] all due respect, no idea what they're talking about... then I have to segue into the \say{this is why accessibility is important} conversation. And we have to start from the total basics, which is... that cuts down on productivity, costs the company money, etcetera, etcetera.} (P7)
\end{quote}

In the face of these options, some BLVSPs chose not to seek formal permission and to simply desist from using their DIY tool: \sayit{[I'd] rather not deal with it [approval process], and not use it [DIY tool]} (P7). On the other hand, to maintain productivity, some participants{\footnote{We omit specific details in the next examples to reduce risk of deanonymization.}} described not notifying employers about their tools or being \sayit{willing to bend the rules, just a little bit}. One participant commented: \sayit{Nah, he [employer] doesn't even know,} while another BLVSP candidly admitted: \sayit{the hacker part of me would probably just carry on using it anyway}.

%why do we even need approval in the first place?
Some BLVSPs questioned \textit{why} DIY tools for accessibility required formal review in the first place, when sighted colleagues' DIY tools for efficiency did not. P1 argued: \sayit{I think [my DIY tools] don't need to be [approved]... It's just some scripts that summarize data, or just present things in a different way.} Similarly, P24 insisted: \sayit{Most developers write little scripts to automate something [and don't seek approval]...I don't believe I'm doing anything wrong.}

%\paragraph{Disapproval and Professional Consequences}
Some companies not only disallowed DIY tools but also strictly prohibited open-source software solutions found on platforms like GitHub, along with commercial workarounds. Even basic ATs and add-ons were frequently denied. P9 reported that \sayit{we need security permission to install tools, and they normally will say no to any tool which an individual has built and no security checks have been done on.} P24 highlighted the necessity to sometimes install open-source software screen readers on servers, which is \sayit{just a huge no for a lot of [employers].} 

Disapproval could have severe professional consequences for our participants, such as losing their jobs. P10, who is currently allowed to use DIY tools for work, imagined being \sayit{fired off} without the tools, \sayit{because my speed would have certainly decreased.} P13 expressed resentment over terminated contracts and lost income due to a company's refusal to either address accessibility issues or authorize DIY solutions: \sayit{As long as I'm getting the work done, I don't know why they're not willing to let me come up with my own workarounds, because it's hard, extra work that I shouldn't have to do, and I'm taking the extra time, still willing to do it.} (P13)
%\sayit{I haven't asked...because I'm 99\% sure I know exactly what they're going to say just because of the amount of regulations.}

%Some participants did not notify the Information Security department of their accessibility tools, as they felt that their tools are no different from those written by sighted developers. 
%\begin{quote}
%    \sayit{Most developers write little scripts to automate something, and they're not going to get info sec approval for a 20-line Python script, right?... I don't believe I'm doing anything wrong that's in my contract.} (P24)
%\end{quote}
% Total words 4.4 as of Sep 11 9am: 15464-12354 = 3110 words (use %TC:ignore and %TC:endignore to calculate)

\subsection{Impact of DIY Tools on the BLV Community}
A number of participants shared their DIY tools with the broader BLVSP and sometimes even BLV community, leading to collaborative tool enhancements and farther-reaching advocacy, tinged with both gratification and regret. Deficiencies in current sharing networks revealed a need for better internal and external support for dissemination.

\subsubsection{Sharing DIY Tools}
Participants alluded to the BLV community's \sayit{culture of sharing} (P12) when explaining their drive to share DIY tools, tips, and tricks through internal and external social networks.
Several participants shared their internal DIY tools with other BLV(SP) employees within the company: \sayit{I put it in an internal repository... it's still [there] and some colleagues have used it} (P15). They sometimes received \sayit{tips and tricks} (P1) from other BLV employees at the company and actively wrote to the internal mailing list about workplace accessibility problems to \sayit{collaborate and brainstorm possible solutions} (P12). Sometimes, a useful tool they found in the internal community became \sayit{a part of my tools as well} (P12).

External BLVSP communities that participants engaged with included Program-L---which was described as \sayit{really the biggest one} (P4), PythonVis, AccessComputing, Lime Connect, Programme Nationale (Portuguese), Ciegos Programadores (Portuguese), the NVDA users mailing list, Access India, and blindcoders.com. BLVSPs saw value in having a mailing list specifically tailored to BLVSPs, since \sayit{it doesn't get as cluttered} (P4) as other lists and \sayit{everyone knows at least something about programming} (P4). In cases where participants were unable to get help from  sighted peers, P5 described leveraging external communities as an additional source of support. 

\subsubsection{Collaborating, Advocating, and Offloading}
%Participants reported collaboration as a naturally emergent when looking for DIY tools or sharing them with BLV networks; together, they created solutions, advocated for accessibility, and sometimes even offloaded the labor of creating DIY tools to others. 

%\paragraph{Finding and Collaborating on DIY Tools}
BLVSPs discovered DIY tools through online communities, as they were \sayit{always looking for tools to make my life easier} (P4). P1 explained that web searches often led him to results \sayit{on a GitHub page or a mailing list.} Similarly, P10 highlighted the recommendations from other BLVSPs on Program-L: \sayit{JAWS scripts, also some old scripts for Visual Studio 2008... This DBeaver thing I told you about, it was also recommended by Program-L.} 

It was not uncommon for the participants to then contribute to some of the tools they found, ranging from suggesting issues to address \sayit{so a lot of the time, if there's something that I think is obvious, that an add-on's not doing, I'll submit an issue about it} (P7), to small code fixes, to modifying another BLVSP's tool when \sayit{they aren't kept up to date} (P10).  

Conversely, when they had published tools, some participants also received contributions from others. P7 recounted receiving pull requests for his DIY tools: \sayit{I was just looking through my messages, and sure enough, a guy from three months ago wanted to do something with [DIY tool].} P16 reported another BLVSP creating \sayit{basically a next version} of his DIY tool: \sayit{He uses my [DIY tool] for everything, basically... And he's rewriting some of it... He's built stuff on it. And he's trying to contribute some of it back.}

The online BLVSP community also fostered \sayit{community-based} (P7), purposefully organized DIY-ing, as participants collaborated with others to addressed shared challenges: \sayit{A few of these [DIY tools] were developed because we were complaining about something in our professional lives that didn't work well... then, Remote Support came about} (P7).
%
%

%\paragraph{Offloading through Advocacy}
Meanwhile, participation in these efforts was not always limited to BLVSPs. Indeed, participants described reducing the collective labor of DIY-ing by coordinating  accessibility advocacy on mainstream sharing platforms. For example, sometimes BLVSPs would share a DIY proof of concept to nudge sighted developers to integrate accessibility features into typical software:
\begin{quote}
    \sayit{%They're [DIY Tools] publicly available on GitHub, and not just to [promote] themselves, but also when advocating to different companies. 
    There's a certain inaccessible tool where [BLVSPs] will just come up with an accessible alternative, and sometimes companies just basically say that \say{This cannot be made accessible; it's just not possible,} which is not true, at all. Then [BLVSPs] will just write prototypes to prove that this is actually doable.} (P13)
\end{quote}
In other cases, BLVSPs strategically sidestepped DIY labor altogether, by rallying around accessibility feature requests:
\begin{quote}
    \sayit{Jupyter Notebook was not being accessible. Through Program-L, we all got together, and one of us decided to open up a GitHub issue, and then everybody would contribute to it, so that it would stay near the top, and be active. And [sighted] developers would see it. And eventually, [they] ... replied to the issue, and now it's a lot more accessible.} (P13)
\end{quote}

\subsubsection{Uptake of DIY Tools}
%Participants' DIY tools were often embraced by other BLVSPs and BLV individuals and further enhanced through mutual contribution. However,  uptake was sometimes bittersweet.

%\paragraph{Pride Through Impact}
BLVSPs took pride in the success and impact of their DIY tools on the BLV(SP) community. Some tools \sayit{went viral} (P24) through the community and were used by many other BLV individuals: \sayit{I just woke up one morning, and it had like 30-something stars on GitHub} (P7). P13 described: \sayit{It's pretty exhilarating... also very satisfying when other people also find it helpful,} while P7 had a \sayit{huge learning experience} as his tool was used in ways he had not anticipated. %Some tools \sayit{went viral} (P24) through the community and were used by many other BLV individuals: \sayit{I just woke up one morning, and it had like 30-something stars on GitHub} (P7). 

Constructive, sometimes \sayit{overwhelmingly positive} (P7) feedback from the community motivated BLVSPs to improve their DIY tools, as others were now able to \sayit{integrate [DIY tool] with their existing workflow} (P7). P13 recounted a time where \sayit{people were able to suggest more things to add,} which helped him extend the tool's capability and \sayit{just have one script that would work with three different platforms.}

%\paragraph{Mutual Contribution to Each Other's DIY Tools}
%Many participants further enhanced DIY tools by contributing to each other's DIY tools, which ranged anywhere from suggesting minimal lines of code fixes to modifying another BLVSP's tool when \sayit{they aren't kept up to date} (P10). Some just simply desired to do so, or because they found it \sayit{easier} (P10) to adapt and maintain existing DIY tools rather than building from scratch. P7 recounted receiving pull requests for his DIY tools” \sayit{I was just looking through my messages, and sure enough, a guy from three months ago wanted to do something with [DIY tool].} P16 reported another BLVSP creating \sayit{basically a next version} of his DIY tool: \sayit{He uses my [DIY tool] for everything, basically... And he's rewriting some of it... He's built stuff on it. And he's trying to contribute some of it back.}

%\paragraph{Bittersweetness of Uptake}
Sometimes, BLVSPs' DIY tools even became inspiration for commercial implementations by companies. One participant\textsuperscript{1} shared how a company implemented and released a feature of a paid product, which \sayit{feels incredibly similar} to his \textit{free}, open-source DIY tool: \sayit{[Company] that made [product]... I'm pretty sure they took a lot of inspiration [for the feature] from [my DIY tool]. Even some of the wording is literally exactly the same as what I wrote.}
%repeats example above and is not worth the word count, IMO -SB

%Another participant was also made aware of the similarities between DIY tool  and the company's feature. Shortly after the participant became aware of the company's feature, \sayit{someone messages me, \say{Have you ever heard of this product called [product]?} Basically, a stand-alone application doing exactly what this thing[DIY tool] did.}

Although tools becoming integrated into mainstream technologies was \sayit{cool, because now more people get access to it,} it left the BLVSP with a bittersweet reminder that their valuable labor was uncompensated and unacknowledged: \sayit{I did it first, just saying. But that's a sign of success, right?}

\subsubsection{Need for Internal Community}
%\paragraph{Valuing and Desiring Internal Mailing Lists}
Many BLVSPs highly valued internal mailing lists as \sayit{one of the best places to get [work-]specific help} (P1): \sayit{For more specific enterprise questions, internal [list] tends to be much more effective than external} (P12). This was particularly because \sayit{[company] has a bunch of technology that only exists at [company]} (P25) and they could not discuss this with external BLVSPs.

%BLVSPs desired access to internal resources like responsive troubleshooting, mailing lists, and BLV employee communities. Yet, only 40\% of our participants' organizations (n = 11) had internal mailing lists for BLV employees. While some participants were unsure whether the mailing list existed, saying, \sayit{maybe there is something like that, and it's just not well-publicized, at all} (P7), others were keenly aware of their absence: \sayit{Honestly, there is nothing that I can reach out to internally} (P5). 

Yet, only 40\% of our participants' organizations (n = 11) had internal mailing lists for BLV employees: \sayit{Honestly, there is nothing that I can reach out to internally} (P5). Most BLVSPs who did not have internal mailing lists strongly desired them, saying that it \sayit{would be extremely valuable} (P7) for sharing \sayit{different tricks and hacks and stuff that would be useful to get your work done, since you guys are all in the same company, completing similar work, or working towards the same goal} (P23). P20 projected that such internal communities would \sayit{be great to build confidence in yourself, \say{Oh, there's somebody working in a similar field as me in the company.}} Moreover, they might \sayit{give everybody a bigger voice} (P25), especially when pushing for accessibility fixes in inaccessible internal tools to ultimately influence company culture: \sayit{If we are able to collaborate and bring out issues collectively... It would help us push for better accessibility as a group and may well open doors for more visually impaired issues down the line.} (P9)

%P29, a senior BLVSP, expressed willingness to support other BLVSPs through such a community:
%\begin{quote}
    %\sayit{I did it the hard way. I'm not happy about the fact that I did it the hard way... It's just I didn't see any %other options... I would engage in the internal mailing list, absolutely... If there are ways I could help others, I %would do that.} (P29)
%\end{quote}

In addition to wanting a mailing list, participants emphasized the need for the company to publicize its existence, making it easier to find. This was partly because they did not want to disclose their disability while seeking such a community. P18 explained, \sayit{I definitely felt like sometimes, some colleagues judged... so I was hesitant to even talk to some people,} fearing that and that disclosure to seek community may \sayit{bite me later} in performance reviews. P7, who was unsure whether community existed at his company, argued for active publicizing, \sayit{so that people do know that they have that resource to take advantage of} and not \sayit{suffer in silence.}

\subsubsection{Need for Centralized External Community}
%Participants highlighted the necessity for \textit{external} BLV(SP) communities to complement the internal. Some BLVSPs worked in companies that are not \sayit{big enough to have multiple people with disabilities} (P15) to form an internal mailing list, resulting in him \sayit{hav[ing] to find the answer outside of my company.} However, participants noted the lack of a centralized, mainstream, and usable platforms.

%\paragraph{Lack of Mainstream BLV(SP) Community}
BLVSPs expressed how they \sayit{definitely need a place where we can find each other} (P22), and desired for a centralized platform for BLVSPs to share and discover DIY tools and knowledge. Participants lamented the dissolution of Twitter (currently X), which used to be \sayit{the place} (P7) for BLVSPs to connect and share resources.  Now, the BLV(SP) community is \sayit{very much scattered} (P7), \sayit{segregated} (P16) across many different platforms, and therefore \sayit{hard to find} (P22). For example, while Program-L is \sayit{probably the largest resource that's out there} (P17), three participants \sayit{didn't know they existed} (P29). %P22 was particularly affected by this void as he made the transition from being a sighted developer to a blind developer.

Difficulties finding community led to difficulties finding and sharing tools. P24 described how \sayit{it's harder to find tools now}, requiring \sayit{either painstaking research, having a pint with other blind people, mailing lists, and Reddit, I guess. It's pretty sad} (P16). On the flip side, some tools were unable to be shared as BLVSPs \sayit{didn't know where to share... there's not a centralized place to share JAWS scripts} (P10). 

In addition to having a place for all their tools, participants wanted a specific, dedicated community where members have a \sayit{shared understanding} (P7) and could answer \sayit{not just programming questions... [but] visual impairment related programming questions} (P13), along with strategies and tools to overcome accessibility barriers at work. They also wanted \sayit{a booklet or public repositories} (P10) that documented \sayit{some common tips and tricks for this kind of stuff... Like a Wiki... a place where people are encouraged to contribute and just describe, \say{this is how a blind person may interact with a diff tool}} (P1). P16 added how newcomers to the BLVSP identity should be equipped with a handbook: \sayit{someone goes blind on Tuesday, and then 10 years later they find out that there's 10 mailing lists [for BLVSPs]. People need to be handed a book of all the things that exist for blind people on day one} (P16).

%Participants lamented the dissolution of the former mainstream BLVSP community, Twitter (currently X). P7 noted how Twitter used to be \textit{the place} to connect and share resources. They found value in the former mainstream (i.e., including blind and sighted people, developers and non-developers), centralized community \sayit{that bound subcommunities together} (P24). Unfortunately, BLV users left the platform because it became \sayit{a whole lot less accessible now that it's been taken over} (P24), and \sayit{there's nothing that has replaced Twitter yet} (P24). As a result, 

%P24 described how \sayit{it's harder to find tools now} (P24) now that the centralized community had vanished. As such, the current way to find out useful DIY tools is through \sayit{either painstaking research, having a pint with other blind people, mailing lists, and Reddit, I guess. It's pretty sad} (P16). 

Participants differentiated sharing on mainstream platforms with relatively less BLV engagement (e.g., GitHub) versus lesser known, highly specialized communities (e.g., Program-L), and each comes with its own limitations. Mainstream platforms have a large audience but may hinder discoverability of \say{blindness tools} (P24): 
\begin{quote}
    \sayit{There's a website called Hacker News, which is for techy people, and that will occasionally recommend tools to me, which is good, but they're not blindness tools... The blindness tools are never going to get promoted on that [website], so I definitely feel like I've lost something now that I don't get those recommendations [from Twitter].} (P24)
\end{quote}
Specialized communities, on the other hand, are harder to locate, if they even exist (e.g., P10 described the absence of a JAWS scripts platform, and P7 noted that the NVDA Add-On store was only recently created), and reach fewer people. Participants described the email-based nature of Program-L as \sayit{overwhelming} (P7): \sayit{It's really ridiculous that mailing lists are the main way} (P16). %These tools are also less likely to be optimized for tool sharing. The email-based nature of Program-L could sometimes get \sayit{overwhelming};  \sayit{Work-wise I probably get about 200 emails every day across different projects I'm on. I try to keep on top of it, but when I add mailing lists into the list, it's just a lot} (P7). Participants described the email-based nature of Program-L as \sayit{overwhelming} (P7): that it agreed, saying \sayit{It's really ridiculous that mailing lists are the main way.} 

%this is good stuff, but it's too much and not in the right order to tell the story, IMO. So, I'm commenting it out wholsale and integrating some of the pieces into surrounding paragraphs -SB

\section{Discussion}

HCI researchers have been investigating how BLV individuals employ life hacks to address inaccessibility in their everyday lives~\cite{herskovitz_hacking_2023}, and Software Engineering researchers have studied how programmers create customized DIY solutions to fit their work needs~\cite{smith_diy_2015}. To the best of our knowledge, this is the first study to explore the experience of BLVSPs, who have dual identities of (1) the BLV hacker and (2) the software professional hacker, about their experiences in building and using DIY solutions for the workplace. 

%Below, we discuss the ``Double Hacker Dilemma'' of DIY-ing for the workplace as a BLVSP. Our findings suggest the need for better company support for building and using DIY solutions in the workplace, as well as a centralized platform for BLVSPs to share DIY tools and knowledge, recommendations for which are presented below.

\subsection{The Double Hacker Dilemma}

The type of hacking we found BLVSPs engaging in stems from the fact they exhibited two intersecting identities: one as a software professional hacker and the other one as a BLV hacker. BLVSPs, just like other software professionals, built their own tools to streamline tasks at work ~\cite{amini_creative_2024, liu_creative_2002, groeneveld_creative_2022}. In the process, they equally acted as creative problem solvers and enjoyed coding solutions to address problems and needs. Moreover, in line with prior research on work hacking~\cite{jetha_smarts_2019, stein_leadership_2022} and software developer DIY-ing~\cite{smith_diy_2015}, our participants were motivated by the desire to be more efficient and help others, and their tools served purposes like general automation and monitoring.

At the same time, however, our participants' BLV identities necessitated hacking not just for convenience, but for overcoming accessibility barriers. While much research has observed such hacking outside the workplace~\cite{buehler_thingverse_2015, hurst_empowering_2011, das_weaving_2020, momotaz_usage_2023}, our study shows that 
%due to their BLV identity, 
BLVSPs had critical additional motivations to DIY at work. This included wanting to be independent and seeking greater efficiency so as to shave off the ``accessibility tax'' of being a screen reader user--all in service of 
%such motivations have been described by prior research on DIY activities of BLV individuals, albeit outside the workplace setting~\cite{buehler_thingverse_2015, hurst_empowering_2011, das_weaving_2020, momotaz_usage_2023}. 
%We also identify motivations specific to BLVSPs, such as 
maintaining their sense of dignity and a professional image while staying apace with sighted colleagues.

%BLVSPs were ``double hackers,'' being both (1) BLV hackers and (2) software professional hackers. Like other software professionals~\cite{amini_creative_2024, liu_creative_2002, groeneveld_creative_2022} who possess the software professional hacker identity, BLVSPs were creative problem solvers, often enjoying coding solutions to address problems and needs. In line with prior research on work hacking~\cite{jetha_smarts_2019, stein_leadership_2022} and programmers DIY-ing~\cite{smith_diy_2015}, our participants' motivations to build DIY tools included wanting to be more efficient and help others, and DIY tools served purposes like general automation and monitoring. However, due to their BLV identity, we found that BLVSPs had additional motivations to DIY, such as wanting to be independent, in control, and seeking greater efficiency so as to shave off the ``accessibility tax'' of being a screen reader user; such motivations have been described by prior research on DIY activities of BLV individuals, albeit outside the workplace setting~\cite{buehler_thingverse_2015, hurst_empowering_2011, das_weaving_2020, momotaz_usage_2023}. We also identify motivations specific to BLVSPs, such as wanting to maintain their professional image and staying apace with sighted colleagues at work.

Building DIY tools for workplace accessibility was a double-edged sword for our participants. Through building DIY tools, participants not only made their work lives better, but at times were recognized by management and in some cases even earned promotions. Yet, this did not necessarily offset the intense labor of struggling through inaccessible, clunky tools, trying to use existing workarounds and failing, building DIY solutions, and going through release, maintenance, and corporate compliance processes. Here, we see how BLVSPs' visible work on the job was only made possible by the uncompensated, invisible access labor~\cite{pandey_understanding_2021, cha_participate_2024} they did behind the scenes.

To capture the tension that exists behind this double-edged sword, we introduce the notion of the ``Double Hacker Dilemma'' to explain the difficult situation that BLVSPs repeatedly face: request organizational troubleshooting by submitting accessibility support tickets that may never be addressed, or take matters into their own hands and leverage their expertise to build DIY solutions? This dilemma is uniquely experienced by BLVSPs. It is distinct from the situation for non-disabled software professional hackers, because they can choose not to DIY, yet still perform their work as normally without `standing out' as incapable amidst their co-workers. It is equally distinct from the situation for non-software professional BLV workers, who, lacking the technical expertise to create solutions, necessarily are dependent on others (co-workers, organizational troubleshooting, Program-L advice) to provide solutions. BLVSPs live at the intersection: they have the ability to solve their own problems, but should they?
%and is distinct from the situation for non-disabled software professional hackers, who, if they choose not to DIY can still perform their work as normally and not `stand out' as incapable amidst their co-workers. We found that BLVSP, however, do not have this luxury of choosing not to DIY, especially if their company did not provide timely, adequate solutions.
%and BLV hackers, who typically possess only one of the two hacker identities. 

When our participants submitted support tickets as BLV employees, requests for support typically languished in the backlog for months, sometimes never being resolved. Our BLVSPs, equipped with their additional software professional hacker identities, thus often chose to 
resolve the problem by leveraging their expertise and DIY-ing solutions such as 
the command line interface tool connecting to the code review system and identifies the code changes for review or the accessible, light version of Microsoft SQL Manager, to name two examples among the many in \autoref{tab:tools}. Doing so, however, typically and unfortunately came at the expense of uncompensated labor, time, and effort. Further compounding the dilemma, even when they built a successful solution, external companies could reject their fix or internal compliance barriers could prevent use within their workplace. 

Our participants' experiences challenged the notion of DIY as a leisure activity~\cite{Davies_hacking_2018}. Individuals---including our BLVSPs---found hacking enjoyable and personally fulfilling~\cite{Davies_hacking_2018, herskovitz_hacking_2023}, and sometimes found that it enhanced task productivity~\cite{McEwan_smart_2016, bloom_hacking_2021, jetha_smarts_2019, stein_leadership_2022}. However, DIY-ing out of necessity to address accessibility shortcomings at work led to these positive feelings frequently being usurped by a sense of resentment being token solution providers, having to spend personal time, and their tools being singled out and sanctioned, compared to their sighted colleagues’ DIY tools that were built for higher productivity and efficiency. The double hacker dilemma thus adds to the well-known systemic accessibility and accommodation issues across the tech industry~\cite{hill_accommodation_2016}. Even when employers provide BLVSPs with accessible setups, these turn out to be incomplete and not as effective as they could be, which is why our participants rolled up their sleeves to build DIY solutions to address their needs and bring themselves to a more equitable playing field. Two thirds of our participants necessarily engaged in DIY, a number that we conjecture far outpaces that of non-BLVSP colleagues, and is especially poignant because BLVSP DIY-ing concerns essential needs rather than conveniences.

\subsection{Towards a Company Culture of Interdependence}

Our findings revealed contrasting challenges for BLVSPs working in companies of different sizes. In larger corporations, restrictive policies and complex approval processes---a well-known phenomenon in tech companies~\cite{spinellis_organizational_2012, nagy_organizational_2010}---hindered DIY tool use, while infrastructure for internal community support was mostly---though not always---present. In contrast, smaller companies offered flexibility with regards to policies but lacked internal resources to facilitate the discovery and dissemination of DIY tools. With the exception of two participants, BLVSPs did not have access to both a robust internal BLV community and adequate support by the company towards using DIY tools.

Companies are attempting to make their workplaces more accessible~\cite{google_dei_2024, ms_dei_2023}. We argue that this has to go beyond providing accessible technologies, instead requiring a culture shift (in line with, e.g., ~\cite{stone_embedding_2016, schur_corporate_2005}). Some of this culture shift needs to happen organizationally, especially in larger companies that typically are not as nimble as startups in adopting new tools~\cite{wintersgill_law_2024, spinellis_organizational_2012, nagy_organizational_2010, almeida_investigating_2019}. Important first steps include differentiating approval processes for DIY tools requested by BLVSPs from other tools that developers ask to bring into the workplace, streamlining the approval process, procurement mandating accessibility in tools being acquired, and creating affinity or resource groups~\cite{glassman_use_2017}.

We also contend that, in addition to procedural changes, this cultural shift must include changes in colleagues' understanding and attitudes towards BLVSPs and accessible (DIY) tools. Indeed, efficient authorization of DIY tools does not necessarily translate to seamless integration into workflows. For example, the new tool may require sighted colleagues to alter their individual and collaborative work processes. A select few of our BLVSPs were successful in bringing their sighted colleagues into the conversation. As those colleagues come to understand access needs of BLVSPs, they may join them in adopting, advocating for, or even collaborating on developing accessible internal DIY tools. When coworkers acknowledge and enact this interdependence, the culture becomes one that can disrupt ability-based hierarchies by revealing the contributions of people with disabilities~\cite{branham_interdependence_2018}.

\subsection{Towards a Centralized Platform for Sharing DIY Tools and Knowledge}

%While some BLVSPs had access to internal resources, others were unsure whether they existed, and many participants who did not have internal communities for BLV peer support desired one. Confirming prior work on knowledge sharing of PWD on social media ~\cite{sweet_community_2020}, our participants were able to gain access to technologies and advocate for accessibility. We extend existing literature by observing how PWD independently support each other on a non-social-media platform: companies' internal communities. Internal communities of BLV employees supported each other through collaboratively devising DIY solutions and strategies, and gaining access to DIY tools built for inaccessible internal tools. 

%interdependent support—a framework introduced by Branham et al.~\cite{branham_interdependence_2018}—by 

Our findings point to a need for a global, accessible platform for sharing, finding, discussing, and improving DIY tools outside of the workplace. Participants expressed frustration about the currently fragmented BLVSP community following the disintegration of Twitter that once served as a centralized hub for BLVSPs where they could at least discover and share tools. Today, DIY tools are scattered across different subcommunities across many mailing lists, subreddits, GitHub repositories, and Discord servers. This led to a lack of awareness of available tools (aligning with Momotaz et al.'s observation~\cite{momotaz_usage_2023}), participants looking for certain tools but not finding them, and multiple participants building DIY tools serving the same purpose.

%. BLVSPs would not have spent extra time and labor DIY-ing, had tools been more accessible and distributed through the community. 

%Without such platforms, the hurdle for BLVSPs to tap into and extract value generated from within their own community increases significantly. For example, 

The desire for a platform was not only about the DIY tools, however. BLVSPs overwhelmingly expressed a need for a support network beyond their company's BLVSP community, if their company even had one. Especially for those at smaller companies, but often equally so for BLVSPs at larger companies, they wished to 
%discover and disseminate tools, 
meet fellow BLVSPs, feel a sense of belonging and validation, share knowledge~\cite{sweet_community_2020}, and have the ability to actively discuss, make suggestions for, and in some cases actively contribute to the ecosystem of DIY tools for work. For companies that commercially develop ATs and software tools, such a platform can equally help them connect to the BLVSP community, understand its needs, and thereby co-create improved accessible technology~\cite{amann_community_2017}.

Ultimately, a platform such as described here could bring together and seamlessly integrate the best of existing platforms such as GitHub, Twitter (X), Reddit, and Program-L, but re-imagined in a highly accessible manner. For example, the platform could include a searchable listing of available DIY tools and accessibility ratings provide by other BLV users. It may additionally provide a space for BLVSPs to find, communicate with, and provide mutual support (e.g., mentorship) for one another. Today, individual companies mostly benefit from the DIY labor of BLVSPs. The fruits of BLVSPs' largely uncompensated labor, however, ought to be returned to the greater BLV(SP) community and a global DIY-centered platform could make progress toward that vision. 

\section{Limitations and Future Work}

Our sample size (30) is more than double the median(13) reported in accessibility research focusing on BLV individuals~\cite{Mack2021}. However, most of our participants are located in the United States of America (18), with other participants in Europe (6), India (4), and Brazil (2); therefore, our findings do not represent the global experiences of BLVSPs. In addition, our participants come from various countries with different cultures and legislations around disability and accommodations. For instance, the Americans with Disabilities Act (ADA)~\cite{ADA}, a landmark legislation in the United States, was enacted in 1990, while a comparable law in Brazil~\cite{brazil_quotas} was only recently passed in 2015. We did not collect data regarding how different disability legislations and policies in each country shape BLVSPs' experiences DIY-ing for workplace accessibility.

Our participants reported diverse visual abilities. 60\% of participants (n =1 8) were totally or completely blind with little or no light perception, and 30\% (n = 12) had varying degrees of vision. However, we acknowledge that this sample distribution may not fully represent real-world demographics.

The gender demographics of our participants were skewed towards men (26 men, 3 women, 1 non-binary), which is a recognized limitation in software engineering research~\cite{Vasilescu2015, Kohl2022, Outao2024}. Furthermore, we did not explore how intersecting identities, such as race, ethnicity, other disabilities, or sexual orientation, may impact participants' work experiences.

This study did not cover every role in the software development process. However, we gathered experiences and perspectives of a much wider range of job positions in software development compared to many pieces of prior literature. Additionally, our study made sure to include BLVSPs with a wide range of experience, seniority, and positions.

%How various social and legistlative settings could impact workplace accessibility of BLVSPs was outside the scope of our research. However, this area is a warranted direction for future work, as the state of accessibility law and advocacy and culture are different between countries. For example, the ADA that dictates accessibility was passed in 1990 in the United States, but a similar law was only recently passed (2015) in Brazil.

Future work should follow up on the assorted suggestions we make, firstly in terms of investigating how to best design and institute effective, interdependent corporate practices. Based on our findings, future work could also investigate features that BLVSPs desire in a centralized, accessible community (e.g., the list of available DIY tools and their descriptions, an ability to request features, integration with GitHub, success stories around introducing tools): toward this, we encourage designing the community platform with BLVSP users to best implement, deploy, and maintain an effective global BLVSP community platform. Lastly, how different social (e.g., advocacy, culture) and legistlative settings around disability across countries impact workplace accessibility is a warranted direction for future work.

%For example, the ADA that dictates accessibility was passed in 1990 in the United States, but a similar law was only recently passed (2015) in Brazil.

\section{Conclusion}

In this study, we presented our findings from semi-structured interviews with 30 BLVSPs about building and using DIY tools to support accessible software development. We identified four novel themes, each with numerous insights underneath: (1) the DIY tools they built and used, (2) motivations for building and using DIY tools in the workplace, (3) impacts of DIY tools on BLVSPs' work, and (4) impacts of DIY tools on the broader BLV(SP) community. BLVSPs found, built, and maintained a wide range of DIY tools they used daily. Yet, the ``Double Hacker Dilemma'' they faced each time they considered DIY-ing for workplace accessibility highlights the need for companies to significantly improve how they support and value BLVSPs' DIY work. Moreover, the lack of an adequate global BLVSP DIY platform means dissemination and discourse are meager, and notably hinders the emergence of a thriving community. 

\begin{acks}
We would like to thank our participants for their involvement in our study and for providing valuable insights and perspectives. This work was supported by the National Science Foundation (NSF) awards \#2211790, \#2326023, \#2210812, and \#2326489. Also supporting this work are research grants by CNPq Brazil, and Google Award for Inclusion Research.
\end{acks}

\bibliographystyle{ACM-Reference-Format}
\bibliography{manuscript}

%\appendix
\end{document}